# Coherent transfer matrix analysis of the transmission spectra of Rydberg excitons in Cuprous Oxide

Heinrich Stolz[1], Rico Schwartz[1], Julian Heckötter[2], Marc Aßmann[2], Dirk Semkat[3], Sjard O. Krüger[2], Manfred Bayer[1]

[1]Institut für Physik, Universität Rostock, Albert-Einstein-Str. 24, 18051 Rostock, Germany,
[2]Institut für Physik, Ernst-Moritz-Arndt-Universität Greifswald, Felix-Hausdorff-Str. 6, 17489 Greifswald, Germany,
[3]Experimentelle Physik 2, Technische Universität Dortmund, D-44221 Dortmund, Germany

**Abstract**

In this study we analyze the transmission spectrum of a thin plate of Cuprous Oxide in the range of the absorption of the yellow exciton states with the coherent transfer matrix method. We demonstrate that, in contrast to the usual analysis using Beer's law, which turns out to be a rather good approximation only in the spectral region of high principal quantum numbers, a consistent quantitative description over the whole spectral range under consideration is possible. This leads to new and more accurate parameters not only for the Rydberg exciton states, but also for the strengths of indirect transitions. Furthermore, the results have consequences on the determination of the density of electron-hole pairs after optical excitation.

## I. Introduction

The recent experimental observation of excitons with high quantum numbers up to n=25 in cuprous oxide $Cu_2O$ at low temperatures [1] has triggered a series of papers on the behavior of these Rydberg excitons in electric and magnetic fields, as maser materials, for studies of quantum properties of matter etc. (for an overview see [2]). On the other hand we expect that due to their fragile character Rydberg excitons are extremely sensitive probes for every kind of deviation from a perfect crystal, so that the presence of other excitations in the crystal, e.g. a plasma of electron-hole pairs, will influence the absorption lines. Indeed, as has been found recently, the presence of an electron-hole plasma or uncompensated impurities with densities around $10^{10}$ cm$^{-3}$ already quenches the absorption of P states with high quantum numbers (n>20) [3,4,5,5a].

These Rydberg excitons are usually detected as resonances in the transmission spectrum of a thin sample. For details of the experimental setup see [1,2]. In the analysis of the transmission spectra one hitherto assumed [1, 2] that multiple reflections and interferences at the sample surfaces can be neglected due to the high absorption. Then one can convert a measured transmission $T(\lambda_{probe})$ into the absorption coefficient $\alpha(E)$ by the simplified expression

$$\alpha(E) = -\ln(C_{refl} \cdot C_{streu} \cdot T)/d = -\ln\left(\frac{T(\lambda_{probe})}{T_0}\right)/d \,, \quad (1.1)$$

where $C_{refl}$ and $C_{scat}$ are the reflection and scattering losses at the window and sample surfaces, $d$ is the thickness of the sample. Also the use of the optical density $\alpha(E) \cdot d$ is common [1,2]. Since the scattering losses are not well known, one usually looks for a spectral region where due to physical reasons (e.g. in a semiconductor well below the lowest exciton transitions) the absorption coefficient is zero. Then one can take $T_0$ as the measured transmission in this region. In the spectrum shown in Fig. 1 this has been done by starting the measurement well below the yellow 1S ortho-exciton quadrupole transition (marked by *). For a very thin sample also the thickness is not well known. From mechanical measurements we estimate for our sample $d_{Cu_2O} = (33 \pm 3)\mu m$, but one can determine it more accurately by adjusting the absorption coefficient to the well known value around the $\Gamma_3^-$ absorption edge (at $E_{1Sy,\Gamma_3^-} = 2.046165$ eV [6]) given by [7]

$$\alpha(E) = \alpha_0 \sqrt{E - E_{1Sy,\Gamma_3^-}} \text{ meV}^{-1/2}; \quad \alpha_0 = (8 \pm 0.25) \text{ cm}^{-1} \,. \quad (1.2)$$

The dependence of $\alpha$ for higher photon energies is also well known [8,9], so that the indirect absorption coefficient can be continued. Furthermore, it is well-known [10] that the absorption lines of the yellow P states are sitting on top of this phonon background. Due to the fact that both processes are leading to the same final state, we observe for the P lines an asymmetric Lorentzian line shape.

The result for the indirect absorption background is shown by the dashed blue lines in Fig.1. Obviously, there is no agreement between the measured and calculated absorption coefficient around the P lines. The absorption coefficient from Eq. (1.1) is smaller than that expected from the indirect process. On the first sight, this means that the determination of $T_0$ is not correct. However, an additional absorption band must have a very peculiar line shape (it has to exactly compensate for the square root behavior around the threshold) and would also require a much smaller sample thickness. Changing the value of $\alpha_0$ in Eq. (1.2) is not possible, as it is an intrinsic property of $Cu_2O$. Therefore, a highly likely other reason that might lead to such a behavior is an interference effect. Its occurrence seems to be quite possible due to the long coherence length of the single-frequency laser used as light source. Another possibility would be the existence of surface layers, either as a layer of cupric oxide (CuO) due to surface oxidation or as an exciton free dead layer [11].

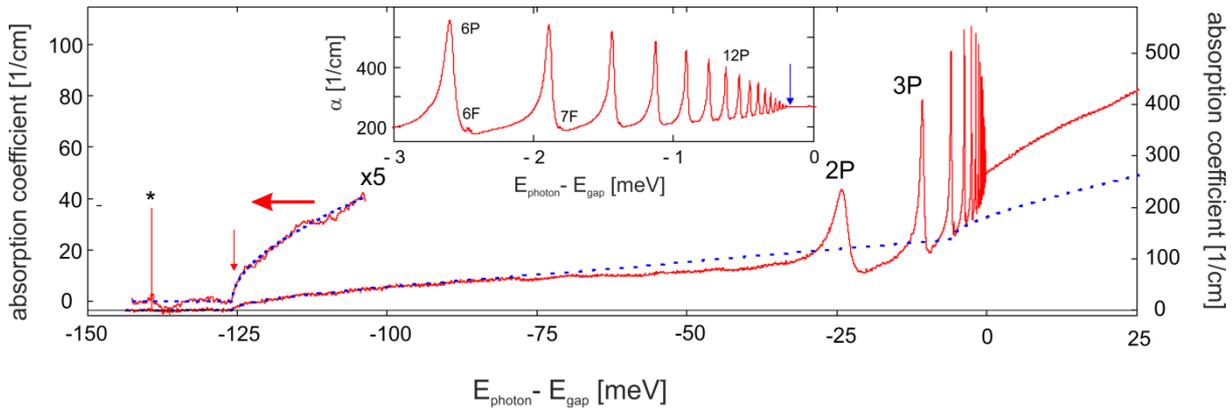

Fig. 1: Experimental results for absorption coefficient of Cu$_2$O (calculated with Eq. (1.1)) from the transmission spectrum (red line) plotted versus the difference of photon energy and band gap from the yellow ortho 1S quadrupole transition (marked by *) to beyond the band gap of the yellow exciton (zero of the abscissa). The left side shows the spectrum around the indirect band gap with enlarged scale (x5). The blue lines are the calculated phonon-assisted absorption. The red vertical arrow indicates the indirect $\Gamma_3^-$ absorption edge. The inset shows the absorption below the band gap showing P lines up to n=22. The blue vertical arrow denotes the apparent gap position.

In this paper, we therefore analyze the transmission spectrum of a thin plate of $Cu_2O$ shown in Fig. 1 taking full account of the coherence of the measuring laser beam by using the well-established coherence transfer matrix (CTM) [12] method to calculate the transmission spectra directly from the dielectric function of the crystal, whereby a single layer of $Cu_2O$ without any exciton dead layer or other surface films is sufficient. Furthermore, we consider the optical properties of the whole set-up including reflection and scattering of the quartz windows of the cryostat and also possible scattering from the optical surfaces of the sample. Adjusting also the strength of the indirect absorption process into the green exciton states, which is not so well known, it is possible to achieve excellent agreement between experiment and theory.

## II. Transfer Matrix calculation of the transmission spectrum

In deriving the dielectric response of the excitons from the transmission spectrum by the CTM method, which is our final goal, we face the serious problem that despite the long coherence length of the probe laser, we did not observe interference effects in the experimentally measured spectrum (see Fig. 1). The reason why no interference effects are observed may be twofold. On the one hand, the sample may show a slight wedge. Here it turns out that already a small wedge of the order of $w = 1\,\mu\text{m/mm}$ is enough to completely remove any interference effect by incoherently averaging the transmission over the finite laser spot size ( $FWHM = 170\,\mu\text{m}$ ). The other possibility would be a certain surface roughness, which also reduces the interference effects (see e.g. [13]). In the following analysis we include both effects. Therefore, we describe the light propagation through the sample by the coherent transfer and propagation matrices method (CTM), whereby the transfer matrices are modified due to interface scattering [13]

$$\mathbf{H}_{ij} = \frac{1}{\gamma\,\tau_{ij}} \begin{pmatrix} \gamma^2 \tau_{12}\tau_{21} + \alpha\,\beta\,\rho_{21}^{\,2} & \beta\,\rho_{ij} \\ \alpha\,\rho_{ij} & 1 \end{pmatrix}. \qquad (1.3)$$

The reflection and transmission coefficients for perpendicular incidence are given by

$$\tau_{ij} = \frac{2\tilde{n}_i}{\tilde{n}_i + \tilde{n}_j} \quad \rho_{ij} = \frac{\tilde{n}_i - \tilde{n}_j}{\tilde{n}_i + \tilde{n}_j} , \tag{1.4}$$

and the roughness parameters are defined as

$$\alpha = e^{-2\left(\frac{2\pi s_{ij} n_i}{\lambda}\right)^2}, \quad \beta = e^{-2\left(\frac{2\pi s_{ij} n_j}{\lambda}\right)^2}, \quad \gamma = e^{-\frac{1}{2}\left(\frac{2\pi s_{ij}}{\lambda}\right)^2 (n_j - n_i)^2}, \tag{1.5}$$

with a surface or interface roughness of RMS deviation $s_{ij}$. Note that already a roughness of the order of $\lambda/10$ reduces the coherence effects significantly.

The propagation matrix is

$$\mathbf{L}_j = \begin{pmatrix} e^{-i\beta_j} & 0 \\ 0 & e^{i\beta_j} \end{pmatrix} \quad \beta_j = \frac{2\pi}{\lambda_0}(n_j - i\kappa_j)d_j \tag{1.6}$$

with $d_j$ the thickness and $\tilde{n}_j = n_j - i\kappa_j$ the complex index of refraction of layer j.

For a single plate of $Cu_2O$ the system matrix would be (with 1 denoting the outside vacuum and 2 the sample)

$$\mathbf{S} = \mathbf{H}_{12}\mathbf{L}_2\mathbf{H}_{21} . \tag{1.7}$$

The reflection and transmission coefficients are obtained as

$$\rho = \frac{S_{12}}{S_{22}} , \quad \tau = \frac{1}{S_{22}} , \tag{1.8}$$

from which we obtain the reflectivity R and transmission T by taking the modulus square, as the medium on both sides is vacuum. These expressions can be easily extended to more complicated layer structures.

Since the diameter of the probe beam is finite, we have to average over the thickness variation across the probe beam. Assuming a Gaussian intensity distribution $\sim \exp(-\rho^2/2\sigma_L^2)$ and a wedge $w$ we have a thickness distribution $\sim \exp(-\rho^2/2\sigma_d^2)$ with $\sigma_d = w\sigma_L$. The effects of the 6 cryostat windows can be taken into account by normalizing the transmission by the reflection and scattering losses $((1-R_Q(\lambda))(1-S_Q(\lambda))^{12}$, where $R_Q(\lambda)$ and $S_Q(\lambda)$ are the reflection and scattering of a single surface (note that due to the window thickness of several mm we can here neglect the coherence of the light beam).

To obtain the optical parameters of the P excitons like oscillator strength, linewidth and resonance energies, we have to calculate the full dielectric function (DF) $\varepsilon(\omega) = \varepsilon_r(\omega) + i\varepsilon_i(\omega)$ from which the complex index of refraction

$$\tilde{n} = n - i\kappa, \tag{1.9}$$

can be calculated as

$$n(\omega) = \sqrt{\frac{1}{2}\left(\varepsilon_r(\omega) + \sqrt{\varepsilon_r(\omega)^2 + \varepsilon_i(\omega)^2}\right)},$$

$$\kappa(\omega) = \sqrt{\frac{1}{2}\left(-\varepsilon_r(\omega) + \sqrt{\varepsilon_r(\omega)^2 + \varepsilon_i(\omega)^2}\right)}, \tag{1.10}$$

$$\alpha(\omega) = \frac{2\omega}{c_0}\kappa(\omega),$$

the last line giving the absorption coefficient. The real and imaginary part of the DF are related by a Kramers-Kronig transformation

$$\varepsilon_r(\omega) = n_b^2(\omega) + \frac{1}{\pi}\wp\int_{-\infty}^{\infty}\frac{1}{\omega'-\omega}\varepsilon_i(\omega)d\omega,$$

$$\varepsilon_i(\omega) = -\frac{1}{\pi}\wp\int_{-\infty}^{\infty}\frac{1}{\omega'-\omega}\varepsilon_r(\omega)d\omega. \tag{1.11}$$

From Eq. (1.10) it is clear that we need the expression for the dielectric function that includes all excitonic resonances, i.e. the indirect processes involving the various phonons (see [3]), the P absorption lines and the corresponding continuum (see [14,3]) and also a background due to the higher excitonic transitions.

The calculation of the contribution of the phonon background to the imaginary part of epsilon is exemplified for the dominant contribution, the $\Gamma_3^-$ band starting at $E_{1Sy,\Gamma_3^-} = (2.032665 + 0.0135)$ eV [6]. It is given by [8]

$$\varepsilon_{i,ind}(\hbar\omega) = a_{ind}\sqrt{\hbar\omega - E_{1Sy,\Gamma_3^-}}\left(1 + \beta(\hbar\omega - E_{i,\Gamma_3^-})\right)^2 \Phi(\hbar\omega - E_{1Sy,\Gamma_3^-}) \approx$$
$$a_{ind}\sqrt{\hbar\omega - E_{1Sy,\Gamma_3^-}}\left(1 + 2\beta(\hbar\omega - E_{i,\Gamma_3^-})\right)\Phi(\hbar\omega - E_{1Sy,\Gamma_3^-}) \tag{1.12}$$

where $a_{ind}$ is a parameter that has to be adjusted by matching the absorption at energies near the onset to the empirical law Eq. (1.2). It turns out to be $a_{ind} = 7.20663 \cdot 10^{-6} / \mu eV^{1/2}$. The parameter $\beta$ describes the non-linearity due to a wave vector dependent deformation potential [8]. It has been determined to be $0.9106 \cdot 10^{-3} / meV$ for the $\Gamma_3^-$ phonon. At photon energies near the blue exciton states (2.6 eV) the indirect band goes over to the direct transition, the effects of which are already included in the background DK. So we have to introduce a cut off $E_0$, for which we assume $E_0 = 2.5$ eV. By a Kramers-Kronig transform we obtain for the contribution of the indirect transitions to the real part of epsilon (calculated with Mathematica) for $\beta = 0$ (here $E_{g,i}$ denotes the threshold energy)

$$\varepsilon_{r,ind}(\hbar\omega) = a_{ind} \begin{cases} -\dfrac{2}{\pi}\left[\sqrt{E_0 - E_{g,i}} - \sqrt{E_{g,i} - \hbar\omega}\arctan\left(\sqrt{\dfrac{E_0 - E_{g,i}}{E_{g,i} - \hbar\omega}}\right)\right] & \text{for } \hbar\omega < E_{g,i} \\ -\dfrac{2}{\pi}\left[\sqrt{E_0 - E_{g,i}} - \sqrt{\hbar\omega - E_{g,i}}\,\text{atanh}\left(\sqrt{\dfrac{\hbar\omega - E_{g,i}}{E_0 - E_{g,i}}}\right)\right] & \text{for } \hbar\omega > E_{g,i} \end{cases} \quad (1.13)$$

and for the beta-term

$$\varepsilon_{r,ind,\beta}(\hbar\omega) = a_{ind}\cdot\beta \begin{cases} \dfrac{2\sqrt{E_0 - E_{g,i}}(E_0 - 4E_{g,i} + 3\hbar\omega) + 6(E_{g,i} - \hbar\omega)^{3/2}\,\text{atan}\left[\sqrt{\dfrac{E_0 - E_{g,i}}{E_{g,i} - \omega}}\right]}{3\pi} & \text{for } \hbar\omega < E_{g,i} \\ \dfrac{2\sqrt{E_0 - E_{g,i}}(E_0 - 4E_{g,i} + 3\hbar\omega) + 3(\hbar\omega - E_{g,i})^{3/2}\ln\left[\dfrac{\sqrt{\dfrac{E_0 - E_{g,i}}{\hbar\omega - E_{g,i}}} - 1}{\sqrt{\dfrac{E_0 - E_{g,i}}{\hbar\omega - E_{g,i}}} + 1}\right]}{3\pi} & \text{for } \hbar\omega > E_{g,i} \end{cases}.$$

(1.14)

Actually, the phonon background consists of several contributions over which we have to sum appropriately [3] (see Table 1).

From semiconductor optics [10,14,15] we get the following expression for the contribution of the P absorption lines to the imaginary part of the DF

$$\varepsilon_{i,P}(\omega) = \sum_{n=2}^{\infty} f_n \frac{1}{\pi} \frac{\Gamma_n + 2A_n(\omega - \omega_n)}{(\omega - \omega_n)^2 + \Gamma_n^2 \cosh^2\left(\dfrac{\omega - \omega_n}{3\Gamma_n}\right)} \quad . \quad (1.15)$$

Here the sum goes over the P states and $f_n$ denotes a constants proportional to the oscillator strengths [16]. The modification of the linewidth parameter in the denominator takes account of the fact that a Lorentzian line shape should turn into an exponential Urbach-like tail far from the resonance [14]. The factor 3 has been chosen to obtain the best agreement with the experimental line shape.

The contribution of the P lines to the real part of the DF is approximately given by

$$\varepsilon_{r,P} = \sum_n f_n \frac{1}{\pi} \frac{(\omega - \omega_n) - 2\Gamma_n A_n}{(\omega - \omega_n)^2 + \Gamma_n^2} \quad (1.16)$$

as calculated by Kramers-Kronig transform from Eq. (1.15), but neglecting the cosh function.

For the P continuum absorption we have from [16]

$$\varepsilon_{i,c}(\hbar\omega) = a_c\left(1 + \frac{\hbar\omega - E_g}{Ry}\right)\Phi(\hbar\omega - E_g) \quad (1.17)$$

where $E_g$ is the band gap and Ry is the exciton Rydberg energy.

The corresponding real part is then

$$\varepsilon_{r,c}(\hbar\omega) = a_c \begin{cases} \dfrac{1}{\pi}\left[\dfrac{E_0 - E_g}{Ry} - \left(1 + \dfrac{\hbar\omega - E_g}{Ry}\right)\ln\left(\dfrac{E_g - \hbar\omega}{E_0 - \hbar\omega}\right)\right] & \text{for } \hbar\omega < E_g \\ \dfrac{1}{\pi}\left[\dfrac{E_0 - E_g}{Ry} - \left(1 + \dfrac{\hbar\omega - E_g}{Ry}\right)\ln\left(\dfrac{\hbar\omega - E_g}{E_0 - \hbar\omega}\right)\right] & \text{for } \hbar\omega > E_g \end{cases}. \quad (1.18)$$

We also have to treat the Urbach tail for a complete description [8]. Assuming the following form

$$\varepsilon_{iUT}(\hbar\omega) = a_c e^{\frac{-E_g + \hbar\omega}{E_u}} \Phi(E_g - \hbar\omega) \quad (1.19)$$

we obtain for the real part

$$\varepsilon_{r,UT}(\hbar\omega) = \begin{cases} a_c \dfrac{e^{\frac{-E_g + \hbar\omega}{E_u}} \operatorname{Ei}\left(\dfrac{E_g - \hbar\omega}{E_u}\right)}{\pi} & \hbar\omega < E_g \\ a_c \dfrac{e^{\frac{-E_g + \hbar\omega}{E_u}} \Gamma\left(0, \dfrac{\hbar\omega - E_g}{E_u}\right)}{\pi} & \hbar\omega > E_g \end{cases}, \quad (1.20)$$

where Ei(z) denotes the exponential integral and $\Gamma(a, z)$ the incomplete Gamma function [16] .

In total we get for the imaginary part of the DF

$$\varepsilon_i(\omega) = \sum \varepsilon_{i,ind}(\omega) + \varepsilon_{i,c}(\omega) + \varepsilon_{i,UF}(\omega) + \varepsilon_{i,P}(\omega) \quad (1.21)$$

and for the real part

$$\varepsilon_r(\omega) = \sum \varepsilon_{r,ind}(\omega) + \varepsilon_{r,c}(\omega) + \varepsilon_{r,UF}(\omega) + \varepsilon_{r,P}(\omega) . \quad (1.22)$$

From the DF we calculate with Eq. (1.8) and (1.10) the transmission spectrum $T(\omega) = |\tau(\omega)|^2$. The relevant parameters are obtained by searching for the best fit of the calculated spectrum to the experimental data. In this way effects of spectral changes of the reflectivity of the sample, which were neglected in the previous analysis (1,2,8) but amount up to 10%, are consistently taken into account.

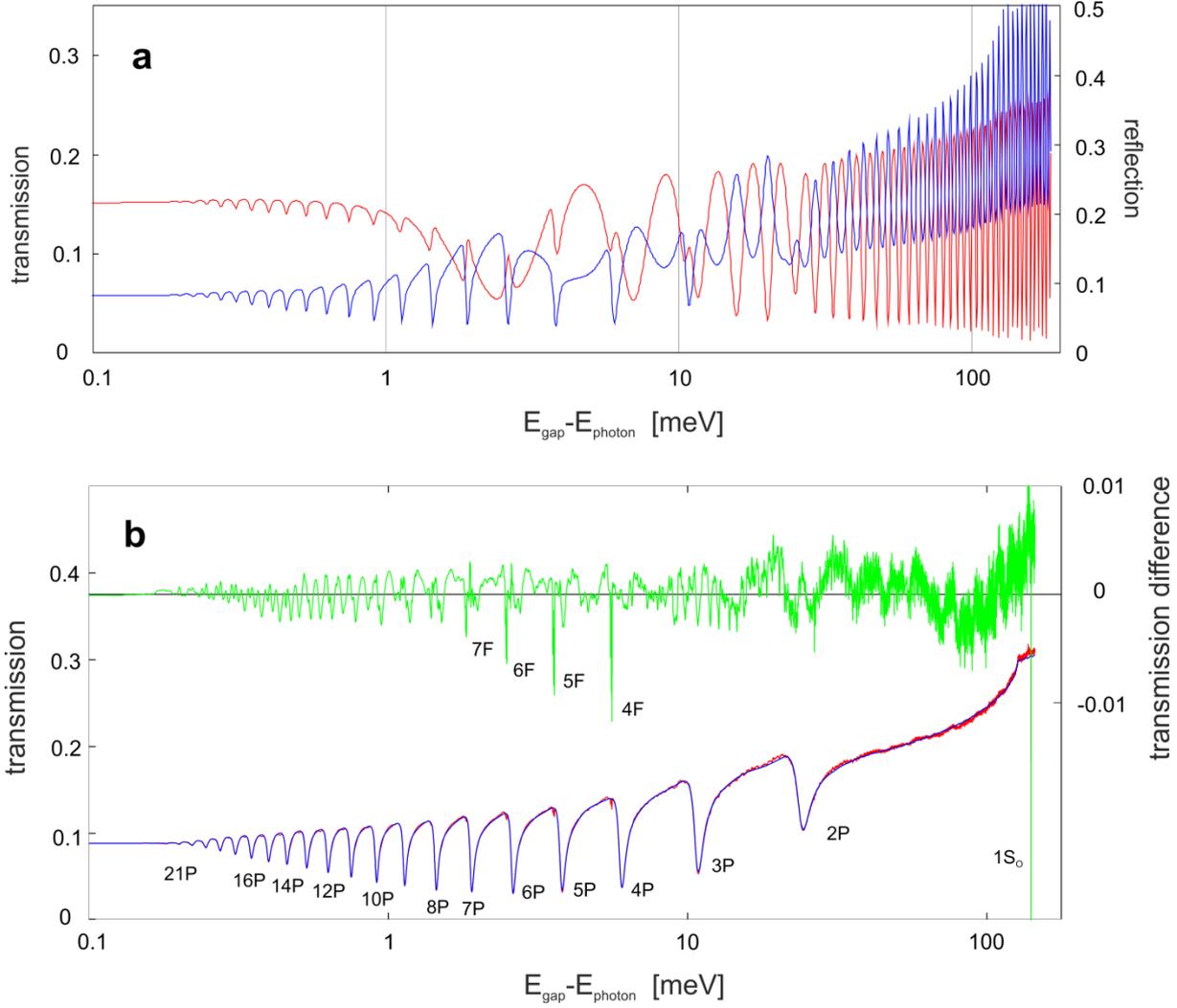

Fig.2, part **a**: Reflection (red) and transmission (blue) spectra of a plate of Cu$_2$O with a constant thickness of $30\,\mu m$, calculated with the coherent transfer matrix (CTM) method. Part **b** shows the comparison of the measured transmission spectrum (red line) with the calculated spectrum assuming a single layer of Cu$_2$O (with parameters given in Tables 1 and 2), the green line gives the difference of the spectra.

To determine the global parameters $s_{Cu_2O}, d_{Cu_2O}$ and $w$, we adjusted the value of the calculated transmission in the spectral region below the indirect absorption threshold $E_{1Sy,\Gamma_3^-}$, where the absorption due to the exciton states should be zero, to the measured one and fitted the transmission in the spectral region close to the threshold to Eq. (1.12).

In Fig. 2a the result of the calculation of the reflection and transmission for a sample with constant thickness is plotted showing pronounced interference fringes. As demonstrated in part b these interference effects are completely absent in a slightly wedged sample ($w = 1$ µm/mm). The comparison with the measured transmission spectrum (red curve in Fig. 2b) shows that a single layer of Cu$_2$O is able to reproduce the whole transmission spectrum from the quadrupole yellow 1S line to the yellow band edge with sufficient accuracy. In comparison of Fig. 2B with Fig. 1 we clearly see that we have to take the coherence properties of the laser into full account in order to describe the transmission spectrum across the whole range from the yellow 1S orthoexciton to the yellow band gap correctly.

The residual differences of the order of a few percent (green curve in Fig. 2b) will be discussed below.

## III. Discussion

The foregoing analysis has shown that it is possible to describe the transmission spectrum fully consistent with the correct dielectric function and the CTM method.

A closer look at the residual differences (green curve in Fig. 2b) reveals that there are four types of deviations behaving quite differently.

1. Most prominent are the sharp peaks slightly above the P exciton lines for n=4 to 8, which are actually triplets. These are the well-known F exciton lines [18], which have not been included in the dielectric function.
2. Above the band gap the transmission of the sample is expected to be smooth, as only continuous processes (indirect phonon-assisted processes and absorption into the continuum) occur. Actually, we observe a pronounced oscillatory pattern shown in Fig. 3a (red line). The oscillation consists of several components with oscillation periods from 100 to 500 µeV. If the origin were Fabry-Perot resonances of various quartz plates, their thicknesses would be between 1.7 and 8 mm, corresponding roughly to the thicknesses of the optical windows. Then we expect these interferences to occur in the whole spectral range with similar periods. Indeed around the indirect absorption edge similar structures do exist (see Fig.3a, blue line).
3. Around the onset of the indirect phonon-assisted transition, we further observe a very broad oscillatory behavior with period around 200 meV. Considering again Fabry-Perot resonances as the potential origin of these oscillations, this would correspond to a quartz plate with thickness of only 4 µm. Because such a thin plate is not present in the setup, one has to consider other origins of this structure. Here one might think of a very thin layer of CuO on the sample which might occur due to oxidation or a thin layer due to polishing. To clarify this further experiments are necessary.

4. Most interesting is the residual transmission spectrum just below the band gap in the region of the Rydberg states (see Fig. 3 b and c). Here it turns out that the weak peaks situated between the P lines show a consistent continuation at the low and high energy side, which, however, scales with the energies of the Rydberg excitons. Therefore,

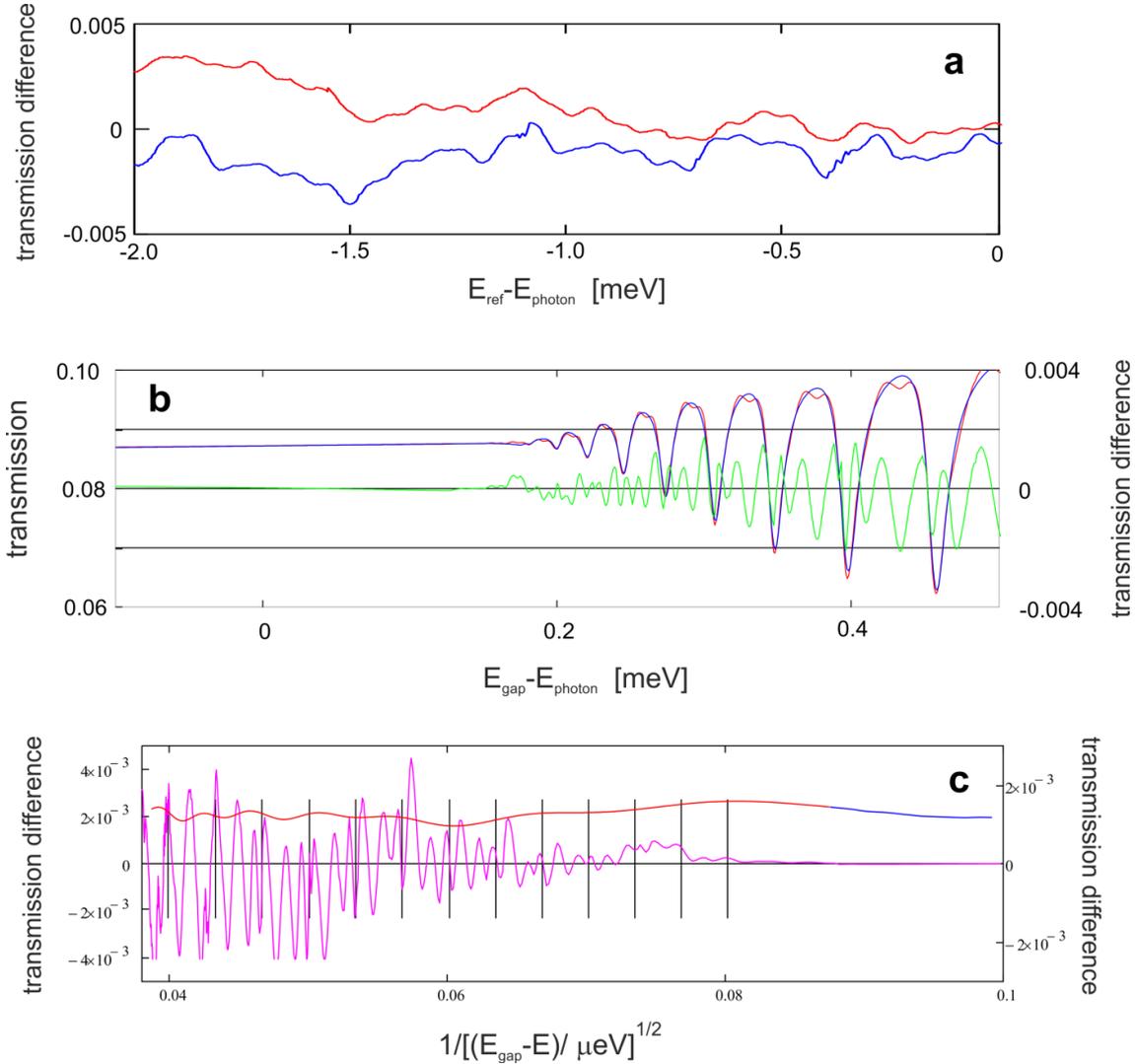

Fig.3: Part a: Comparison of residual transmission spectra directly above the band gap (red curve, $E_{ref} = E_g$) with that 100 meV below the band gap (blue curve, $E_{ref} = E_g - 100$ meV) showing a similar structure. Part b: Transmission spectrum (red experiment, blue calculation) near the band gap showing Rydberg transitions up to n=22. The green colored line gives the difference of measured and calculated spectra showing residual interferences, which are absent above the band gap (arrow). Part c: residual interferences plotted on a scale with equidistant Rydberg states. The red line gives the underground extrapolated to energies below the band gap by the SVD method (see Appendix 1).

these interference structures cannot originate in external effects but must be closely related to the Rydberg states themselves. Obviously, the explanation as a quantum superposition effect of neighboring Rydberg states [19] has to be extended to include the complete manifold. One also might speculate that some kind of polariton effect does lead to these additional interferences like those observed in CdS and other II-VI

11semiconductors [11]. Here further pump-power dependent measurements are necessary.

The parameters of the fit can be grouped into three categories: 1. the parameters of the P lines (resonance energies, oscillator strength, linewidth and asymmetry), 2. the parameters of the continuum absorption (band gap shift, absorption strength and Urbach tail) and of the phonon-assisted indirect transitions, and 3. the parameters describing the sample properties (thickness, wedge, surface roughness). The parameters of the P lines are given in Table 2 and also shown in Fig. 4, the other sets of parameters are given in Table 1.

| parameter | Notation | value | Note | parameter | Notation | value | note |
|---|---|---|---|---|---|---|---|
| thickness | $D_{Cu_2O}$ (μm) | 32.11 | fit | Indirect transitions | $a_{ind}[\mu eV^{-1/2}]$ | $7.20663 \cdot 10^{-6}$ | ref.3 |
| wedge | $w[\mu m/mm]$ | 1 | fit | $1Sy(5+) - \Gamma_3^-$ | A | 1.05771 | ref. 3 |
| | | | | | $E_{1Sy,\Gamma_3^-}$ (meV) | 2046.228 | ref. 9 |
| roughness | $s_{rhg}$ (nm) | 28.7 | fit | $2Sy(5+) - \Gamma_3^-$ | A | 0.75871 | fit |
| | | | | | $E_{2Sy,\Gamma_3^-}$ (meV) | 2151.10 | ref. 22 |
| Band gap shift | $\Delta [\mu eV]$ | -177.7 | fit | $1Sy - \Gamma_4^-(2)$ | | 0.06689 | fit |
| | | | | | $E_{1Sy,\Gamma_4^-(2)}$ (meV) | 2114.828 | ref.22 |
| Strength | $a_c$ | $1.42 \cdot 10^{-3}$ | fit | $1Sg(5^+) - \Gamma_3^-$ | A | 1.99079 | fit |
| | | | | | $E_{1Sg(5+),\Gamma_3^-}$ (meV) | 2167.7 | ref. 22 |
| Urbach1 | $E_{ur1}$ | 142.0 | fit | $1Sg(4^+) - \Gamma_2^-$ | A | 2.74075 | fit |
| | | | | | $E_{1Sg(4+),\Gamma_2^-}$ | 2161.7 | ref. 22 |
| Urbach2 | $E_{ur1}$ | 98.0 | fit | $1Sg(5^+) - \Gamma_5^-$ | A | 0.02640 | fit |
| | | | | | $E_{1Sg(5+),\Gamma_5^-}$ (meV) | 2165.0 | ref. 22 |
| Fraction | $a_{ur}$ | 0.715 | fit | | | | |
| Table 1: values of general parameters used in the fit of the transmission spectrum ||||||||

The resonance energies $E_n$, the oscillator strengths $f_n$, the linewidths $\Gamma_n$ (HWHM!) and the asymmetry parameters $A_n$ are given in Table 2.

Defining a hypothetical "exact" energy of each P state in the form of a modified quantum defect formula [21,22]

$$E_n = -\frac{Ry}{(n-\delta_n)^2} + E_g \text{ with } \delta_n = \delta_0 + \frac{\delta_2}{(n-\delta_0)^2} + \frac{\delta_4}{(n-\delta_0)^4} + \frac{\delta_6}{(n-\delta_0)^6} + \ldots, \quad (1.23)$$

| Quantum number $n =$ | Energy $E_n$ (meV) | Oscillator strength $f_n$ | Linewidth $\Gamma_n$ (µeV) | Asymmetry $A_n$ |
|---|---|---|---|---|
| 2 | -23728 | 15.024 | 1238.9 | -0.450 |
| 3 | -10700 | 9.6731 | 383.5 | -0.320 |
| 4 | -5970.3 | 4.4305 | 140.24 | -0.279 |
| 5 | -3774.6 | 2.3834 | 76.06 | -0.300 |
| 6 | -2592.5 | 1.3714 | 41.85 | -0.301 |
| 7 | -1888.2 | 0.8438 | 27.67 | -0.300 |
| 8 | -1439.0 | 0.5949 | 20.29 | -0.251 |
| 9 | -1129.6 | 0.4013 | 15.81 | -0.250 |
| 10 | -911.14 | 0.2933 | 12.71 | -0.213 |
| 11 | -749.78 | 0.2126 | 10.98 | -0.220 |
| 12 | -628.32 | 0.1612 | 9.83 | -0.221 |
| 13 | -533.53 | 0.1216 | 8.73 | -0.219 |
| 14 | -459.34 | 0.0945 | 7.52 | -0.185 |
| 15 | -399.48 | 0.0784 | 7.14 | -0.161 |
| 16 | -349.94 | 0.0566 | 6.15 | -0.156 |
| 17 | -309.88 | 0.0425 | 5.34 | -0.090 |
| 18 | -276.74 | 0.0329 | 5.1 | 0.0125 |
| 19 | -248.48 | 0.0194 | 5.3 | 0.100 |
| 20 | -223.81 | 0.0115 | 3.3 | 0.125 |
| 21 | -202.18 | 0.0072 | 2.4 | 0.095 |
| 22 | -187.35 | 0.0013 | 3.3 | 0.345 |

Table 2: Results for the parameters of the Rydberg exciton states as obtained from the fits of the transmission spectrum according to the dielectric theory, Eqs. (1.12) - (1.22). The error for the energies and linewidths is about $\pm 1\ \mu eV$, while that for the oscillator strength is about $\pm 0.05$.

the parameters of the best fit are given in Table 3. Equation (1.23) allows us to describe all P lines in the range between n=2 and n=22 with an accuracy of $\pm 1\ \mu eV$ (see Fig. 4a). For the oscillator strength [16] and line width [15] we used a similar quantum defect formula as for the energies

$$f_n = f_0 \frac{(n-\delta_{fn})^2 - 1}{(n-\delta_{fn})^5} \text{ with } \delta_{fn} = \delta_{f0} + \frac{\delta_{f2}}{(n-\delta_{f0})^2} + \frac{\delta_{f4}}{(n-\delta_{f0})^4} + \frac{\delta_{f6}}{(n-\delta_{f0})^6} + \ldots \quad (1.24)$$

| parameter | Energy | parameter | Oscillator strength | parameter | Linewidth |
|---|---|---|---|---|---|
| Ry (meV) | -85.155 | $f_0$ | 237.22809 | $\Gamma_0$ ($\mu eV$) | 8905.9604 |
| $\delta_0$ | 0.2925 | $\delta_{f0}$ | 0.68336 | $\delta_{\Gamma 0}$ | -0.22932 |
| $\delta_2$ | -7.3321 | $\delta_{f2}$ | 5.79769 | $\delta_{\Gamma 2}$ | -3.6534 |
| $\delta_4$ | 3.1544 | $\delta_{f4}$ | -23.20157 | $\delta_{\Gamma 4}$ | -9.6549 |
| $\delta_6$ | -0.5638 | $\delta_{f6}$ | 28.16376 | $\delta_{\Gamma 6}$ | 0 |
| $E_g$ (meV) | 2172.0485 | | | $\Gamma_r$ ($\mu eV$) | 4.23 |

Table 3: Parameters for the Rydberg exciton states according to Eqs. (1.23) - (1.25).

and

$$\Gamma_n = \Gamma_0 \frac{(n-\delta_{\Gamma n})^2 - 1}{(n-\delta_{\Gamma n})^5} + \Gamma_r \text{ with } \delta_{\Gamma n} = \delta_{\Gamma 0} + \frac{\delta_{\Gamma 2}}{(n-\delta_{\Gamma 0})^2} + \frac{\delta_{\Gamma 4}}{(n-\delta_{\Gamma 0})^4} + \frac{\delta_{\Gamma 6}}{(n-\delta_{\Gamma 0})^6} + \ldots \quad (1.25)$$

where $\Gamma_r$ is sample dependent homogeneous broadening constant. For the parameters see Table 3.

The resonance energies of the P excitons and the line widths can be described with Eq. (1.23) or Eq. (1.25) by a generalized quantum defect formula for all quantum numbers from n=2 to 22 with good accuracy (see inset in Fig. 4a). The same holds for the line widths (up to n=20). In contrast, the oscillator strength follows a quantum defect equation only up to n=16, for higher n the effects of charged impurities have to be taken into account [5].

Finally, we discuss the problem of the number of absorbed photons that are converted into elementary excitations in the crystal. Usually, this is calculated from Beer's law as [3]

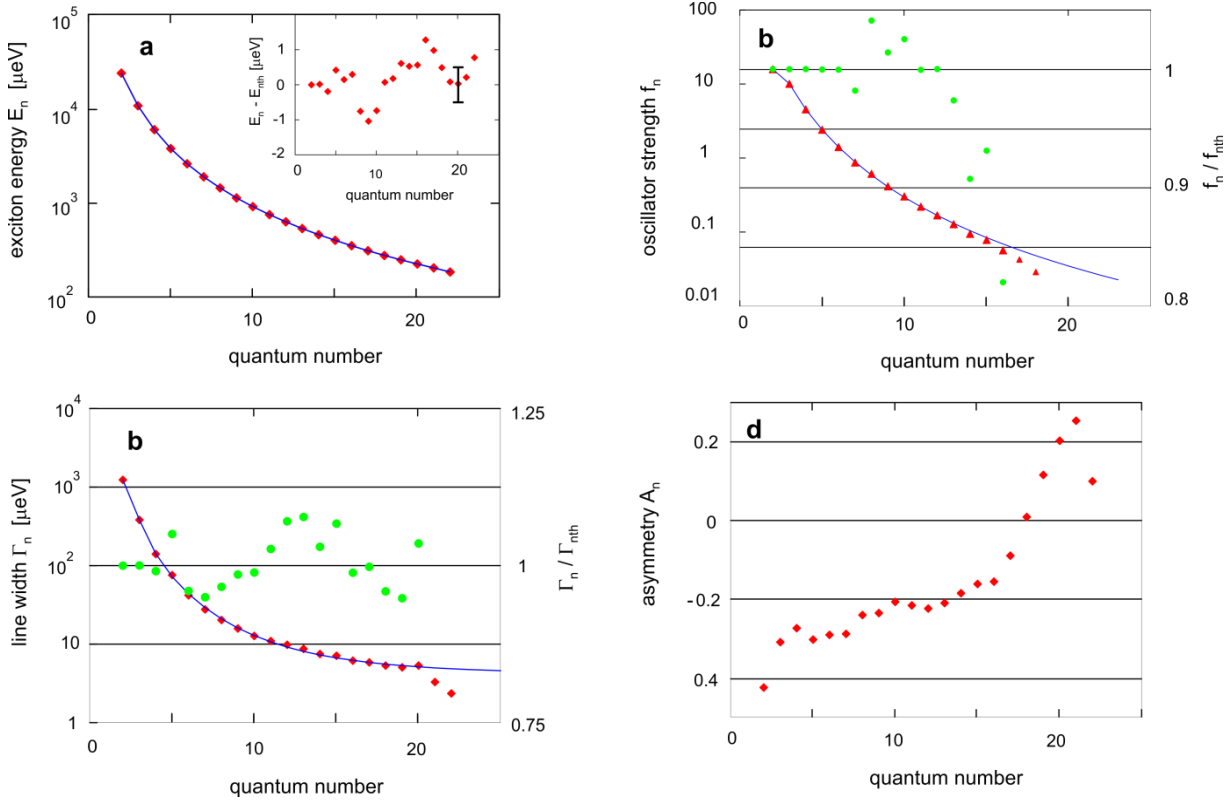

Fig. 4: Energies (a), oscillator strengths (b), linewidths (c) and asymmetry parameters (d) for the Rydberg P exciton states are plotted against the principal quantum number n. The inset in part a gives the difference of the fitted energies to Eq. (1.23). The vertical bar gives the approximate error. The green dots in parts b and c give the ratio of the quantities from the fits to that obtained from Eqs. (1.24) and (1.25) showing that the experimental values are reproduced only up to n=12 for the oscillator strength and to n=20 for the line width.

$$\dot{N}_{phot} = (1 - \exp(-\alpha \cdot d))(1 - R_{Cu_2O}) I_0 \ . \quad (1.26)$$

This, however, is not possible any more, as one has standing waves due to interference inside the crystal and scattering due to surface roughness leading to a loss of laser intensity. Instead one has to calculate the absorbed power per volume of the crystal applying Fermi's golden rule. This leads to the following expression

$$\frac{dP}{dt dV} = 2 \frac{\omega}{n^2} \langle u \rangle \kappa \ , \quad (1.27)$$

where $\langle u \rangle$ is the energy density of the electromagnetic field

$$\langle u \rangle = \frac{1}{2} \varepsilon_0 n^2 |E|^2 \ , \quad (1.28)$$

$n$ and $\kappa$ are the real and imaginary part of the refractive index (see Eq. (1.6)). Here we have assumed that $\kappa \ll n$. The field inside the sample is given by

$$E(z) = E_r \exp(i\beta z/d) + E_l \exp(-i\beta(d-z)/d) \quad (1.29)$$

where $\beta$ is given by Eq. (1.6). $E_r$ and $E_l$ are the electric field strengths of the right and left propagation waves at the surface of the sample at z=0 and can be obtained from the transmission (Eq. (1.8)) by

$$\begin{pmatrix} E_l \\ E_r \end{pmatrix} = \mathbf{L}_1 \mathbf{H}_{12} \begin{pmatrix} 0 \\ \tau \end{pmatrix}. \quad (1.30)$$

This shows that the laser intensity inside the sample might considerably vary spatially over distances of less than 100 nm.

To obtain the total absorbed power inside the sample, one has to integrate over the sample thickness (note that this also requires averaging over the sample surface due to the effects of the wedge). This gives

$$\frac{dP}{dt} = [1 - \exp(-2\omega \cdot \kappa \cdot d / c_0)] \eta_{cor} \cdot I_0 . \quad (1.31)$$

Here $\eta_{cor}$ is given by

$$\eta_{cor} = |E_r|^2 + |E_l|^2 \exp(2\omega \cdot \kappa \cdot d / c_0) + 2\,\mathrm{Re}\left[ E_r \bar{E}_l \exp(i\omega \cdot (n - i\kappa) \cdot d / c_0) \right]. \quad (1.32)$$

As seen from Fig.5, where the results of Eqs. (1.26) and (1.31) are compared, using Beer's law overestimates the absorbed power by almost a factor of two. This has to be taken into account not only in deriving the densities of excitons or electron-hole pairs (see [3]) after optical excitation, but also in the recently published theory of optical nonlinearities of Rydberg excitons for long coherence length of the excitation laser [23]. Also a calculation of the absorbance A by the relation A+R+T=1 does give wrong results (see the corresponding curves in Fig. 5) as the sample shows substantial surface scattering.

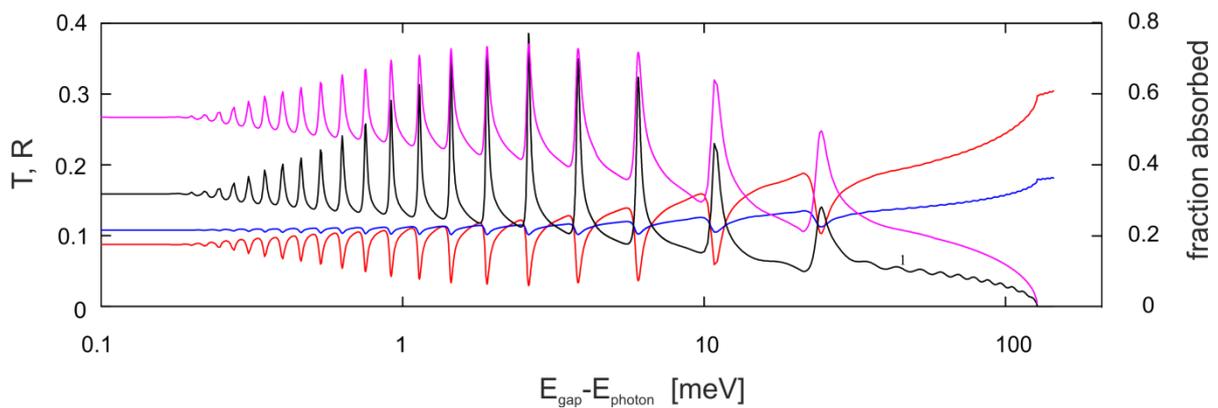

Fig. 5: Transmission (T, red) and reflection (R, blue) spectra (left ordinate) in comparison with the spectrum of the absorbed fraction of laser power incident on the sample using the simple Beer's law (magenta, right ordinate) and the exact calculation in the CTM method (black line, right ordinate).

## VI. Conclusions

In this study we have analyzed the transmission spectrum of a thin plate of $Cu_2O$ in the range of the absorption of the yellow exciton states with the coherent transfer matrix method and demonstrated that a consistent quantitative description is possible. The analysis gives accurate values for the parameters of the Rydberg P excitons such as resonance energies, oscillator strengths and line widths. These can be parameterized by a generalized quantum defect formula. Fortunately, it turns out that the more exact CTM method almost gives the same results for the parameters of the P excitons as the usually used simpler analysis based on Beer's law, at least for high quantum numbers n>10. However, if one needs a highly accurate quantitative analysis, e.g. in the investigations of the action of an electron-hole plasma on the Rydberg states, the CTM method, as exemplified in this paper, must be used.

**Acknowledgements**

D.S. thanks the Deutsche Forschungsgemeinschaft for financial support (project number SE 2885/1-1), the Dortmund side acknowledges the support by the Deutsche Forschungsgemeinschaft through the International Collaborative Research Centre TRR160 (Project A8) and AS 459/1-3.